\documentclass[twocolumn,showpacs,amsfonts]{revtex4-1}
\usepackage{amsmath}
\usepackage{latexsym}
\usepackage{float}
\usepackage{amssymb}
\usepackage{graphicx}
\usepackage{textcomp}
\usepackage{hyperref}
\usepackage{geometry}
\usepackage{color}
 1

\def\ba{\begin{eqnarray}}
\def\ea{\end{eqnarray}}
\def\be{\begin{equation}}
\def\ee{\end{equation}}
\def\bm{\begin{math}}
\def\me{\end{math}}

\newcommand{\dummy}

\begin{document}
\title{Droplet growth during vapor-liquid transition in a 2D Lennard-Jones fluid}
\author{Jiarul Midya and Subir K. Das$^{*}$}
\affiliation{Theoretical Sciences Unit, Jawaharlal Nehru Centre for Advanced Scientific Research,
 Jakkur P.O., Bangalore 560064, India}

\date{\today}

\begin{abstract}
~ Results for the kinetics of vapor-liquid phase transition have been presented from the molecular dynamics simulations of a
single component two-dimensional Lennard-Jones fluid. The phase diagram for the model,
primary prerequisite for this purpose, has been obtained via the 
Monte Carlo simulations. Our focus is on the region very close to the vapor branch of the coexistence curve.
Quenches to such region provide morphology that consists of
disconnected circular clusters in the vapor background. We identified that these clusters exhibit diffusive 
motion and grow via sticky collisions among them. The growth follows power-law behavior
with time, exponent of which is found to be in nice agreement with a theoretical prediction.
\end{abstract} 


\maketitle
\section{Introduction}
~When a homogeneous system is quenched inside the miscibility gap, it falls unstable to fluctuations and 
moves towards the new equilibrium via the formation and growth of particle-rich and particle-poor 
domains \cite{chap3_KBinder1,chap3_AOnuki,chap3_RALJones,chap3_SPuri1,chap3_AJBray}. In addition to being of interest from 
the fundamental scientific point of view, understanding of associated phenomena has many important practical consequences 
\cite{chap3_SPuri1,chap3_Gelb,chap3_Weitkamp,chap3_Quake}, e.g. in designing of advanced materials and devices,
extraction of oil and natural gases, understanding of cloud physics, etc. In the context of phase separation
in solid mixtures, some features of kinetics are well understood
\cite{chap3_KBinder1,chap3_AOnuki,chap3_RALJones,chap3_SPuri1,chap3_AJBray}. However, significant challenges
remain when at least one of the phases is fluid. The objective of the current work is to understand
the dimensionality dependence of kinetics
for phase separation in a vapor-liquid transition, for quenches with low overall density.
\par
The nature of a domain pattern is quantitatively studied via, 
among other quantities, the two-point equal time correlation function, which, in an isotropic situation, 
has the definition \cite{chap3_SPuri1,chap3_AJBray} ($r=|\vec{r}|$)
\begin{eqnarray}\label{chap3_corfn}
 C(r,t) =<\psi(\vec{0},t)\psi(\vec{r},t)>-<\psi(\vec{r},t)>^2,
\end{eqnarray}
where $\psi$ is a space ($\vec{r}$) and time ($t$) dependent order-parameter. The angular brackets in Eq. (\ref{chap3_corfn})
are related to the statistical averaging, involving space and initial configurations.
Typically, during the growth, the structures at different times are self-similar \cite{chap3_SPuri1,chap3_AJBray} (in statistical sense). 
As a consequence, $C(r,t)$ exhibits the scaling property \cite{chap3_SPuri1,chap3_AJBray}
\begin{eqnarray}\label{chap3_scorfn}
 C(r,t)\equiv \tilde{C}(r/\ell).
\end{eqnarray}
In Eq. (\ref{chap3_scorfn}), $\ell$ is the average size of the domains or clusters, which usually exhibits power-law 
growth with time as \cite{chap3_KBinder1,chap3_AJBray,chap3_AOnuki,chap3_SPuri1}
\begin{eqnarray}\label{chap_leng}
\ell \sim t^{\alpha}. 
\end{eqnarray}
The exponent $\alpha$ depends upon the system and order-parameter dimensionality \cite{chap3_SPuri1,chap3_AJBray}, 
transport mechanism 
\cite{chap3_SPuri1,chap3_AJBray,chap3_HTanaka1,chap3_HTanaka2,chap3_IMLifshitz,chap3_KBinder2,chap3_KBinder3,
chap3_EDSiggia,chap3_HFurukawa1,
chap3_HFurukawa2}, order-parameter conservation \cite{chap3_SPuri1,chap3_AJBray}, as well as the type of pattern 
\cite{chap3_KBinder2,chap3_KBinder3,chap3_EDSiggia,chap3_SRoy3}. 
\par
For a vapor-liquid transition, relevant nonequilibrium order-parameter can be constructed from the local 
density field $\rho_{\vec{r}}(t)$. We define $\psi(\vec{r},t)=\rho_{\vec{r}}(t)-c$,
$c$, e.g., can be taken to be the (equilibrium) critical density ($\rho_c$) or
the value of density at the coexistence diameter. In this work we will chose $c=0.35\simeq \rho_c$.
Integration of this scalar order parameter, for the present problem, 
over the whole system remains constant with time. For quenches close to $\rho_c$, say via the 
variation of temperature ($T$), one expects an 
interconnected domain structure \cite{chap3_AJBray,chap3_SPuri1,chap3_SRoy3,chap3_SMajumder2}. 
On the other hand, in the nucleation and growth regime, 
close to the coexistence curve, one of the phases (liquid or vapor) fails to 
percolate \cite{chap3_RShimizu,chap3_Cdatt,chap3_SRoy3,
chap3_KBinder2,chap3_KBinder3}. There has been significant recent interest in the kinetics with
such morphology \cite{chap3_SRoy3,chap3_RShimizu,chap3_Cdatt,chap3_SRoy1,chap3_SRoy2,chap3_SKdas2,
chap3_SMajumder4,chap3_JJung1,chap3_JJung2,chap3_SRazavi,chap3_Warren,chap3_Perrot}.
For overall density close to the vapor branch, circular or spherical liquid droplets, 
depending upon the system dimension ($d$), nucleate \cite{chap3_ACZettlemoyer,chap3_FFAbraham,chap3_KBinder4}. 
Associated problems have direct relevance in the context of 
cloud physics \cite{chap3_STwomey,chap3_EIlotoviz}. 
\par
The above mentioned droplets 
should retain their shape while growing \cite{chap3_SRoy1,chap3_SRoy2,chap3_SRoy3}. During phase separation in 
solid binary mixtures \cite{chap3_SPuri1,chap3_IMLifshitz}, such droplets, formed by the minority particles in 
an asymmetric composition, are practically static \cite{chap3_SMajumder4} and 
growth in the system occurs via an evaporation-condensation mechanism, proposed 
by Lifshitz and Slyozov (LS) \cite{chap3_IMLifshitz}. 
In this mechanism, particles from a smaller droplet get detached, to be diffusively deposited on a 
larger droplet. The value of $\alpha$ in that case is $1/3$, irrespective of the value of $d$. 
In fluids, on the other hand, these droplets are expected to 
have significant mobility \cite{chap3_SRoy3,chap3_KBinder2,chap3_KBinder3}. 
For the diffusive motion of the droplets and coalescence following collisions, the growth law is predicted by 
Binder and Stauffer (BS) \cite{chap3_KBinder2,chap3_KBinder3,chap3_EDSiggia}. For this mechanism, 
solution of the equation \cite{chap3_EDSiggia}
\begin{eqnarray}\label{chap3_bs_law}
\frac{dn}{dt}=-Bn^{2},
\end{eqnarray}
$n$ being the droplet density ($\propto 1/\ell^d$) and $B$ a constant, provides 
\begin{eqnarray}\label{chap3_gexp_alpha}
\alpha=\frac{1}{d}.
\end{eqnarray}
The right side of Eq. (\ref{chap3_bs_law}) is related to the collision frequency. Assuming that the
collisions are sticky, this is equated with $dn/dt$. For the mechanism under discussion, 
$B$ may be a function of both $\ell$ and the droplet diffusivity $D$. This quantity can 
be treated as a constant, accepting the validity of the generalized Stokes-Einstein-Sutherland relation 
\cite{chap3_JPHansen,chap3_TMSquires}. The value of $B$, however, may have dependence upon 
temperature \cite{chap3_JPHansen,chap3_TMSquires}. This will modify the growth amplitude, depending upon the depth of quench.
\par
Though the growth exponent in Eq. (\ref{chap3_gexp_alpha}) was predicted for the liquid-liquid transitions, it was 
recently shown \cite{chap3_SRoy3,chap3_SRoy1,chap3_SRoy2}, from studies in $d=3$, that even for the vapor-liquid 
transitions this theory works, if the background vapor density is reasonably high with long range fluctuations. 
But in $d=3$, the value of the exponent is same as the LS one. Thus, $d=2$ provides a better ground for the confirmation 
of the mechanism and the validity of Eq. (\ref{chap3_bs_law}) since in this dimension the BS value is different from the LS one.
\par
In this paper we study the kinetics of phase separation in a single component Lennard-Jones (LJ) system, via the molecular 
dynamics (MD) simulations \cite{chap3_DFrankel,chap3_MPAllen} in $d=2$. For a very low overall density, at 
reasonably high temperatures, we observe nucleation and growth of circular liquid droplets in the vapor background. 
Via the calculation of the mean-squared-displacements (MSD) of the centers of mass (CM) of the droplets we confirmed 
their diffusive motion. It has been shown that between collisions, the change in the number of particles in a 
droplet is negligible, implying growth via the 
inter-droplet collisions. Finally, the exponent $\alpha=1/2$, as predicted by BS \cite{chap3_KBinder2,chap3_KBinder3}, 
is observed. For choosing a region of interest 
inside the miscibility gap, reasonable knowledge of the coexistence curve becomes essential. 
This we have obtained via the Monte Carlo simulations \cite{chap3_DPLandau}.
\section{Model and Methods}
~ In our model, particles $i$ and $j$, at a distance $r$ from each other, interact via \cite{chap3_MPAllen}
\begin{eqnarray}\label{chap3_mod_LJ}
 U(r)=u(r)-u(r_c)-(r-r_c)\frac{du}{dr}\Big|_{r=r_c},
\end{eqnarray}
where $u(r)$ is the standard LJ potential \cite{chap3_MPAllen}
\begin{eqnarray}\label{chap3_LJ}
 u(r)=4\varepsilon \Big[\Big(\frac{\sigma}{r}\Big)^{12}-\Big(\frac{\sigma}{r}\Big)^{6}\Big],
\end{eqnarray}
$\sigma$ being the particle diameter and $\varepsilon$ the interaction strength. The cut-off distance 
$r_c$ ($=2.5\sigma$), in Eq. (\ref{chap3_mod_LJ}), was introduced to facilitate faster computation. 
The discontinuity in the force, thus appears, was taken care of by the introduction of the last term in Eq. (\ref{chap3_mod_LJ}).
\par
~Phase diagrams for single component LJ systems in $d=2$ were previously calculated by other 
researchers \cite{chap3_MRovere,chap3_ADBruce}. In these works, however, the potentials were slightly different from ours. 
These authors either did not use a cut-off or did not introduce a force correction term after the truncation. 
While MD simulation with the 
full LJ potential is extremely time consuming, a truncated one without force correction 
is not recommended for such simulations, since jumps in energy, pressure (P), etc. may occur. 
The modification of the potential \cite{chap3_MPAllen} in our work was made by keeping these problems in mind. 
However, the last term in Eq. (\ref{chap3_mod_LJ}) 
reduces the nearest neighbor energy. Thus, it is expected that the phase diagram for our model will be 
different from those in Refs. \cite{chap3_MRovere,chap3_ADBruce}. It becomes then necessary 
to obtain at least a working phase diagram for the present model.
\par
We estimated the phase behavior, including the critical values for temperature ($T_c$) and density, via the 
Gibbs ensemble Monte Carlo (GEMC) simulation method \cite{chap3_DFrankel,chap3_AZPanagiotopoulos,chap3_DPLandau}. 
For the kinetics, we have performed MD simulations in the canonical ensemble, using various hydrodynamics preserving 
thermostats \cite{chap3_DFrankel,chap3_EAKoopman,chap3_SDStoyanov,chap3_TSoddemann,chap3_SNose}, 
the results from all of which match with each other. For the sake of convenience, we present results only from the Nos\'{e}-Hoover 
thermostat (NHT) \cite{chap3_DFrankel} which controls the temperature better.
\par
The GEMC simulations \cite{chap3_DFrankel} 
were performed in two boxes, for each combination of $T$ and $\rho$,
the latter being the overall density. There we have allowed three
different types of trial moves, viz., particle displacements in and volume change of each of the boxes, as well as particle transfer
from one box to the other, by keeping the total number ($N$) of the particles and total area ($V$) of the boxes fixed.
We have fixed the overall density $(\rho=N/V)$ to a value $0.3$. This procedure, in the long time limit, leads
to different constant density profiles (with fluctuations around the mean values)
for the two boxes, one corresponding to the vapor phase and the other to the liquid phase,
if the chosen temperature is below the critical value. The phase diagram was obtained by plotting these mean values
as a function of temperature.
\par
All the simulations for kinetics were performed in periodic square boxes with area $V=L^2$, $L$ ($=2048$ in units of $\sigma$) 
being the linear system dimension. Other than the snapshots, the results are presented after averaging over at least $15$ 
independent initial configurations. Unless otherwise mentioned, for all the simulations, we fixed $\rho$ 
to $0.03$ and $T$ to $0.35\varepsilon/k_B$. Time in our MD simulations was measured in units of $\sqrt{m\sigma^2/\varepsilon}$, 
where $m$ is the mass of the particles. From here on, for the sake of convenience, we set $m$, $\sigma$, $\varepsilon$ and $k_B$ to unity.
\par
For the calculation of the $C(r,t)$, used for the verification of the self-similarity property \cite{chap3_AJBray} 
and obtaining the values of $\ell$, we have mapped the continuum systems onto the (square) lattice ones. 
If the density (calculated by considering the nearest neighbors) at a lattice point is higher than the 
critical value, $\psi$ is assigned the value $+1$, otherwise $-1$. We have obtained the values of $\ell$ from 
\begin{eqnarray}\label{chap3_cross_crfn}
 C(\ell,t)=0.25,
\end{eqnarray}
as well as from the first moment of the domain size distribution function, $P(\ell_d,t)$ \cite{chap3_SMajumder2}, as 
\begin{eqnarray}\label{chap3_pdf}
 \ell=\int \ell_d ~ P(\ell_d,t) ~ d\ell_d,
\end{eqnarray}
where $\ell_d$ is the distance between two successive interfaces along any direction. 
We have also calculated the length scale by appropriately identifying the droplets, thus their radii (via
circular structural approximation). Results from these methods are 
proportional to each other. Except for Fig. \ref{fig3_4}, we have used $\ell$ from Eq. (\ref{chap3_pdf}).

\section{Results}
\begin{figure}[h!]
\centering
\includegraphics*[width=0.4\textwidth]{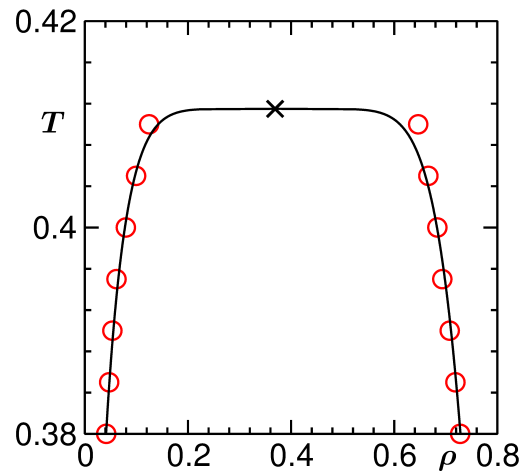}
\caption{\label{fig3_1} Vapor-liquid coexistence curve for the 
considered two-dimensional ($2D$) Lennard-Jones fluid, in temperature vs density plane. 
The circles are from the Gibbs ensemble Monte Carlo simulations and the continuous curve is obtained by fitting the simulation 
data to the theoretical form corresponding to the criticality in the $2D$ Ising model. 
The cross mark is the location of the critical point. For the simulation
results we have used $V=1250$.}
\end{figure}

~ In Fig. \ref{fig3_1} we show the coexistence curve, in $T$ vs $\rho$ plane, for the model system. 
The circles are from the GEMC simulations \cite{chap3_AZPanagiotopoulos}. 
The values of $T_c$ and $\rho_c$ were estimated by fitting the 
simulation data (in the finite-size unaffected region) to the equations ($A$ and $A^{\prime}$ are constants) \cite{chap3_DFrankel}
\begin{equation}\label{chap3_dens_diff}
 \rho_{\ell}-\rho_{v}=A(T_c-T)^{\beta},
\end{equation}
and
\begin{equation}\label{chap3_avg_diff}
\frac{\rho_{\ell}+\rho_{v}}{2}=\rho_{c}+A'(T_c-T),
\end{equation}
where $\rho_{\ell}$ and $\rho_v$ are the densities along the (high density) liquid and the (low density) vapor branches of 
the coexistence curve, respectively. For even better estimation of $\rho_c$, more terms in ($T_c-T$) \cite{chap3_NBWilding,chap3_YCKim}, 
with powers $1-\alpha$, $2\beta$ ($\alpha$, zero for $d=2$ Ising critical universality class, being the critical exponent 
for specific heat), may be needed in Eq. (\ref{chap3_avg_diff}). Since the objective here is not to 
accurately estimate critical singularities, we avoid these terms. We, however, mention that the value of $\rho_c$ will be slightly less if the 
term with exponent $2\beta$ is included in the analysis. In the fitting exercises (using the above equations)
we have set $\beta$ to $1/8$, the $d=2$ Ising critical exponent for the order parameter \cite{chap3_DPLandau}. 
Given the short range of the LJ interaction, it is expected that our model will belong 
to the Ising critical universality. This exercise provides $T_c\simeq 0.41$ and $\rho_c\simeq 0.37$,
represented by the cross. The continuous line in Fig. \ref{fig3_1} represents the Ising behavior, 
with which the simulation data are in nice agreement.
\par
As mentioned above, given that the inter-particle interaction in our model is weaker than those in the previous 
works \cite{chap3_MRovere,chap3_ADBruce}, a smaller value of $T_c$ is expected. The differences of
this value of $T_c$ from those of the other works are not due to significant finite-size effects or any other errors in our analysis.
\par
It is necessary to stay reasonably away from $T_c$ to avoid fluctuations along interfaces 
that may affect the identification of the droplets. Furthermore, as will be elaborated later, the divergence of the 
relaxation time for the density field will bring crucial finite-size effects in the kinetics,
necessitating very long simulations with very large systems, if we choose to be very close to $T_c$.
Keeping these facts in mind, we study kinetics at $T=0.35$.
\par
\begin{figure}[h!]
\centering
\includegraphics*[width=0.4\textwidth]{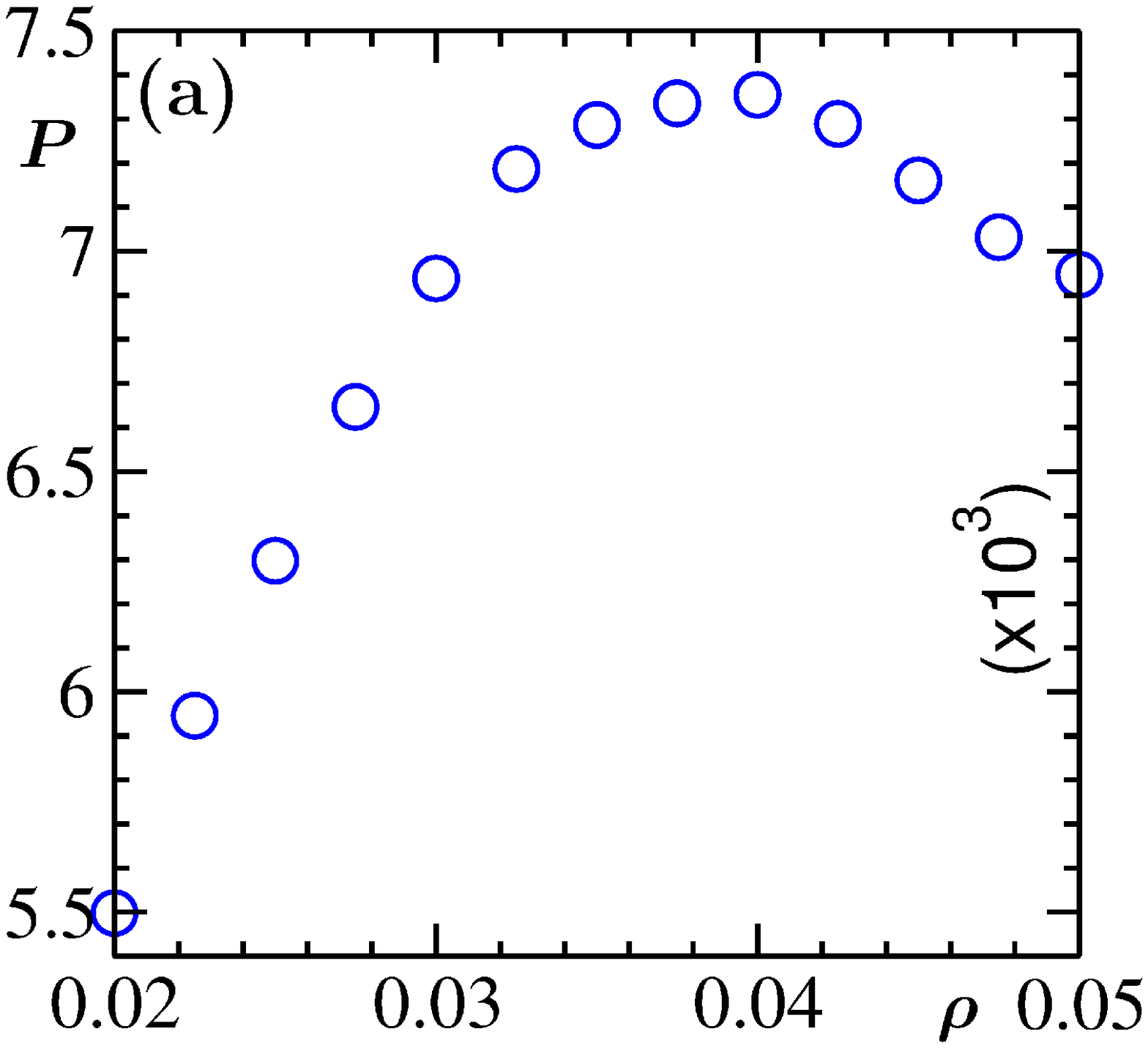}
\includegraphics*[width=0.4\textwidth]{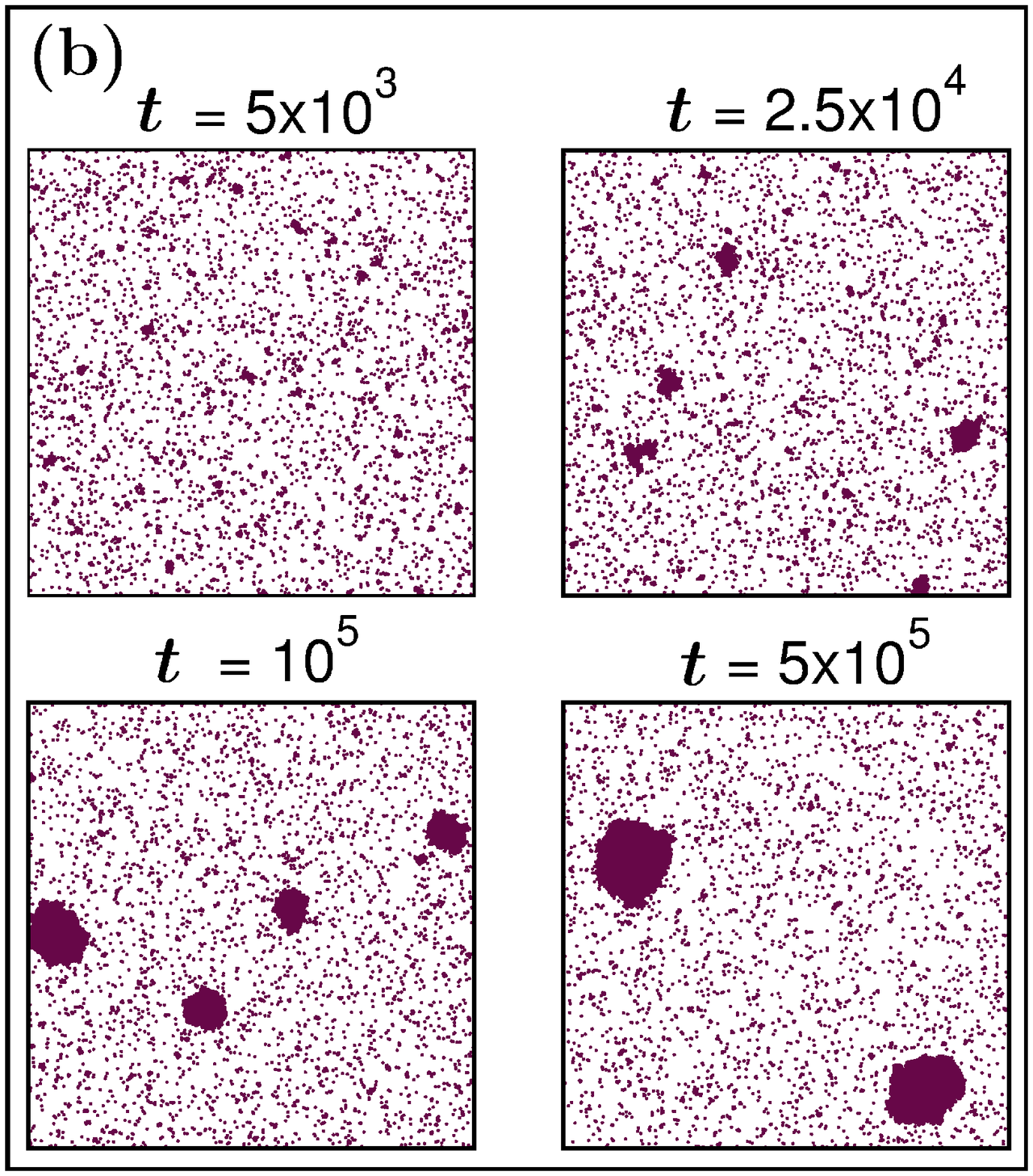}
\caption{\label{fig3_2} (a) Plot of $P$ vs $\rho$ at $T=0.35$. These results, obtained
via MD simulations in square boxes with $L=50$, are presented after averaging over $50$ independent
initial configurations. (b) Snapshots during the evolution of the LJ fluid, having been quenched from a high temperature homogeneous 
state, with overall density $\rho=0.03$, to a temperature $T=0.35$, inside the coexistence curve. The dots mark the location
of the particles. Though the results are obtained for $L=2048$, we have shown only small parts ($400\times400)$ of the original system.}
\end{figure}

Before proceeding to presenting the results for kinetics, we identify a suitable value for density, for the above mentioned 
temperature, so that we observe nucleation and growth of droplets. 
For this purpose, in Fig. \ref{fig3_2}(a) we show a plot of the pressure, as a function 
of density, at $T=0.35$. We have calculated $P$ from the time average of the diagonal elements of the two dimensional stress 
tensor. In the metastable or nucleation regime one expects $\partial P/\partial V <0 $, implying $\partial P/\partial \rho > 0$.
We choose $\rho=0.03$, for which the latter condition is satisfied.
\par
Fig. \ref{fig3_2}(b) shows the evolution snapshots following the quench of a high temperature homogeneous 
system, to the state point of our interest. Here note that, because of the very low overall density, 
nucleation of stable droplets \cite{chap3_ACZettlemoyer,chap3_FFAbraham} requires fluctuations over long distances. 
Such fluctuations are, however, rare. Thus, the onset of phase separation gets delayed. 
This fact can be appreciated from the first snapshot that contains clusters 
of very small sizes, despite the value of $t$ being quite large. Following nucleation, rather fast growth of the disconnected 
clusters is clearly visible. 
\par
The presence of the circular symmetry, as seen in Fig. \ref{fig3_2}, in the structure of the clusters is related to 
the minimization of the interfacial free energy. The (minor) roughness that is noticeable in the boundary regions of these 
droplets, despite being reasonably away from the critical point, can be due to the fact that the line
fluctuations in $d=2$ are larger than the surface fluctuations in $d=3$. At temperatures even closer to $T_c$, 
such interfacial fluctuations \cite{chap3_AOnuki,chap3_VPrivman} will make the identification of the droplets very difficult.
\begin{figure}[h!]
\centering
\includegraphics*[width=0.4\textwidth]{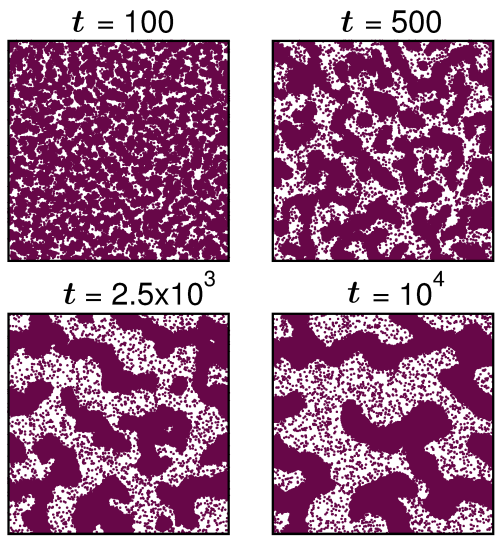}
\caption{\label{fig3_3} 
Same as Fig. \ref{fig3_2}(b) but for overall density $\rho=0.35$. These results correspond to $L=512$, unlike all the other results
related to kinetics (with NHT). Little higher density of the vapor phase that appears here, compared to that
in Fig. \ref{fig3_2}, is because of the fact that in Fig. \ref{fig3_2} we presented $400\times 400$ cuts
(from a larger system).
}
\end{figure}
\par
In Fig. \ref{fig3_3} we present a few evolution snapshots for a high overall density, viz., $\rho=0.35$, the value of $T$ remaining 
the same as in Fig. \ref{fig3_2}. The pattern in this case is contrastingly different
from that in Fig. \ref{fig3_2}; the high density quench provides an elongated, 
interconnected morphology \cite{chap3_SPuri1,chap3_AJBray,chap3_SRoy3,chap3_SMajumder2}. The mechanisms of growth in the two cases are also 
expected to be different \cite{chap3_KBinder2,chap3_KBinder3,chap3_EDSiggia,chap3_HFurukawa1,chap3_HFurukawa2}. However, in the rest of the paper we will focus only on the off-critical quench.
\begin{figure}[h!]
\centering
\includegraphics*[width=0.4\textwidth]{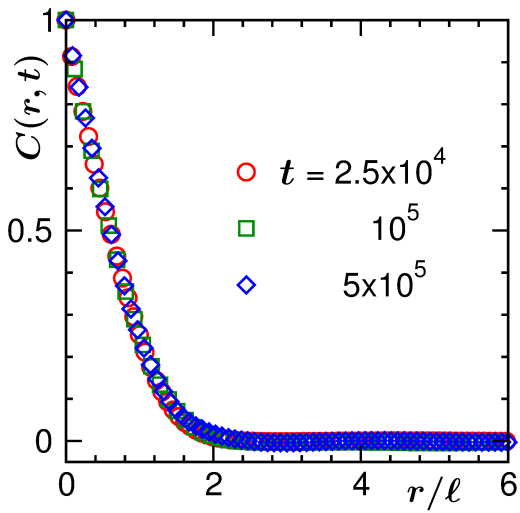}
\caption{\label{fig3_4} Scaling plot of the two-point equal time correlation function. Here we have plotted $C(r,t)$ as a function
of the scaled distance $r/\ell$, using data from three different times.}
\end{figure}
\par
First, to check for the self-similar property \cite{chap3_AJBray} of the 
structures at different times, in Fig. \ref{fig3_4} we show the scaling plot 
of the two-point equal time correlation function. Nice collapse of data from different times, when plotted vs the scaled distance $r/\ell$, 
confirms that the patterns at different times differ from each other only by a change in length scale. Here we mention, for the 
bicontinuous structure (that we observed for $\rho=0.35$), the correlation function exhibits prominent oscillation, 
albeit damped, around zero.
This has connection with the fact that the integration of the $C(r,t)$ over space is related to the total system 
order-parameter \cite{chap3_SPuri1,chap3_AJBray}, the latter 
being approximately zero (in the language of $\psi$) for $\rho$ close to $\rho_c$. Observation of only a very shallow minimum in the present case is 
due to the fact that, for $\rho << \rho_c$, the composition with respect to negative and positive values of $\psi$ is highly asymmetric. 
\begin{figure}[h!]
\centering
\includegraphics*[width=0.4\textwidth]{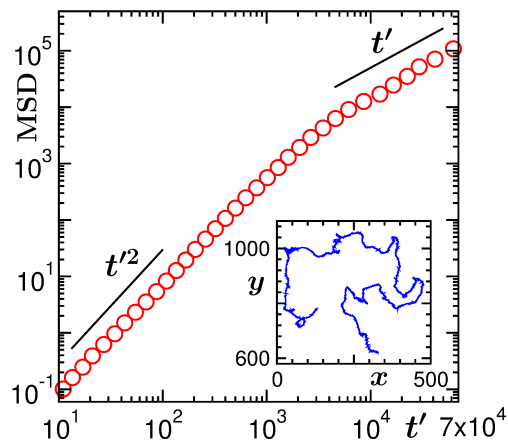}
\caption{\label{fig3_5} Mean-squared-displacement (MSD) for the center of mass of a typical liquid droplet is plotted vs time, 
on a log-log scale. The power-law regimes, parallel to $t'^2$ and $t'$, correspond, respectively, to ballistic and diffusive motions. 
The inset shows the trajectory of the center of mass of the droplet over a period of time.}
\end{figure}
\par
To understand the mechanism of growth, next we calculate the MSD \cite{chap3_JPHansen} 
of the CMs of the droplets as 
\begin{equation}\label{chap3_msd_dif}
 \mbox{MSD}= \Big\langle \Big( \vec{R}_{\mbox{CM}}(t')-\vec{R}_{\mbox{CM}}(0) \Big)^2\Big\rangle,
\end{equation}
where $\vec{R}_{\mbox{CM}}(t')$ is the location of a cluster CM at time $t'$.
For this purpose, the droplets were appropriately identified by using the connectivity of regions with the positive values of $\psi$. 
The MSD for a typical CM is presented in the main frame of Fig. \ref{fig3_5}, as a function of time, on a log-log scale. 
Here $t'$ is not the simulation time, it is measured from the moment a probe starts. 
\par
As mentioned above, the growth of the droplets during kinetics of phase separation in solid binary mixtures occurs 
via the diffusion of particles from smaller droplets to larger ones, the CMs of the droplets remaining essentially fixed. 
However, as expected \cite{chap3_SRoy3,chap3_KBinder2,chap3_KBinder3}, Fig. \ref{fig3_5} shows that the droplets 
can have significant movement in fluids. 
At early time, say upto $t'=100$, the data are reasonably proportional to $t'^2$, implying ballistic motion \cite{chap3_JPHansen}. 
After this time, the data gradually turn over to a linear behavior, that corresponds to diffusive motion \cite{chap3_JPHansen}. 
Such a diffusive motion can be appreciated from the inset of this figure where we show a trajectory of the droplet under consideration. 
\begin{figure}[h!]
\centering
\includegraphics*[width=0.4\textwidth]{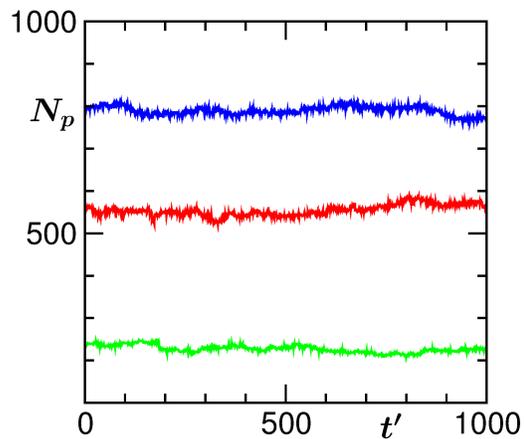}
\caption{\label{fig3_6} Plots of the 
numbers of particles in a few droplets, with the variation of time, the later being calculated from the 
beginning of the probes. During this period the droplets do not undergo collisions with any other droplet.}
\end{figure}
\par
The mobility of the droplets will allow them to collide with each other. We have checked that such collisions are sticky in nature. 
If the droplets are in the liquid phase, mobility of the constituent particles, with respect to the CMs, is rather high. 
This fact allows a noncircular cluster, that has formed after a collision between two droplets, to gain circular shape, that is 
required to minimize the interfacial free energy, before it undergoes a collision. This explains the structural 
self-similarity, thus the scaling property of $C(r,t)$.
\begin{figure}[h!]
\centering
\includegraphics*[width=0.4\textwidth]{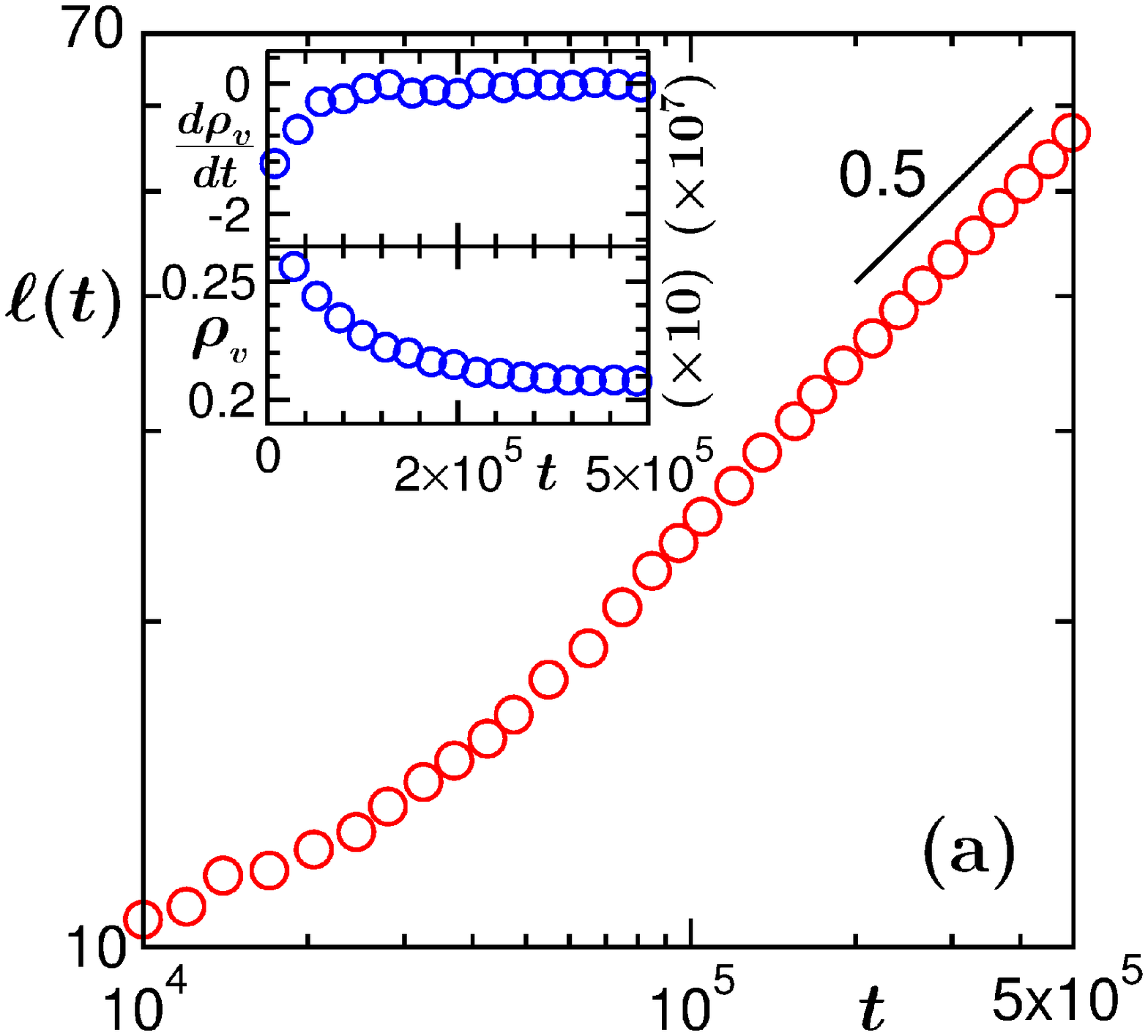}
\includegraphics*[width=0.4\textwidth]{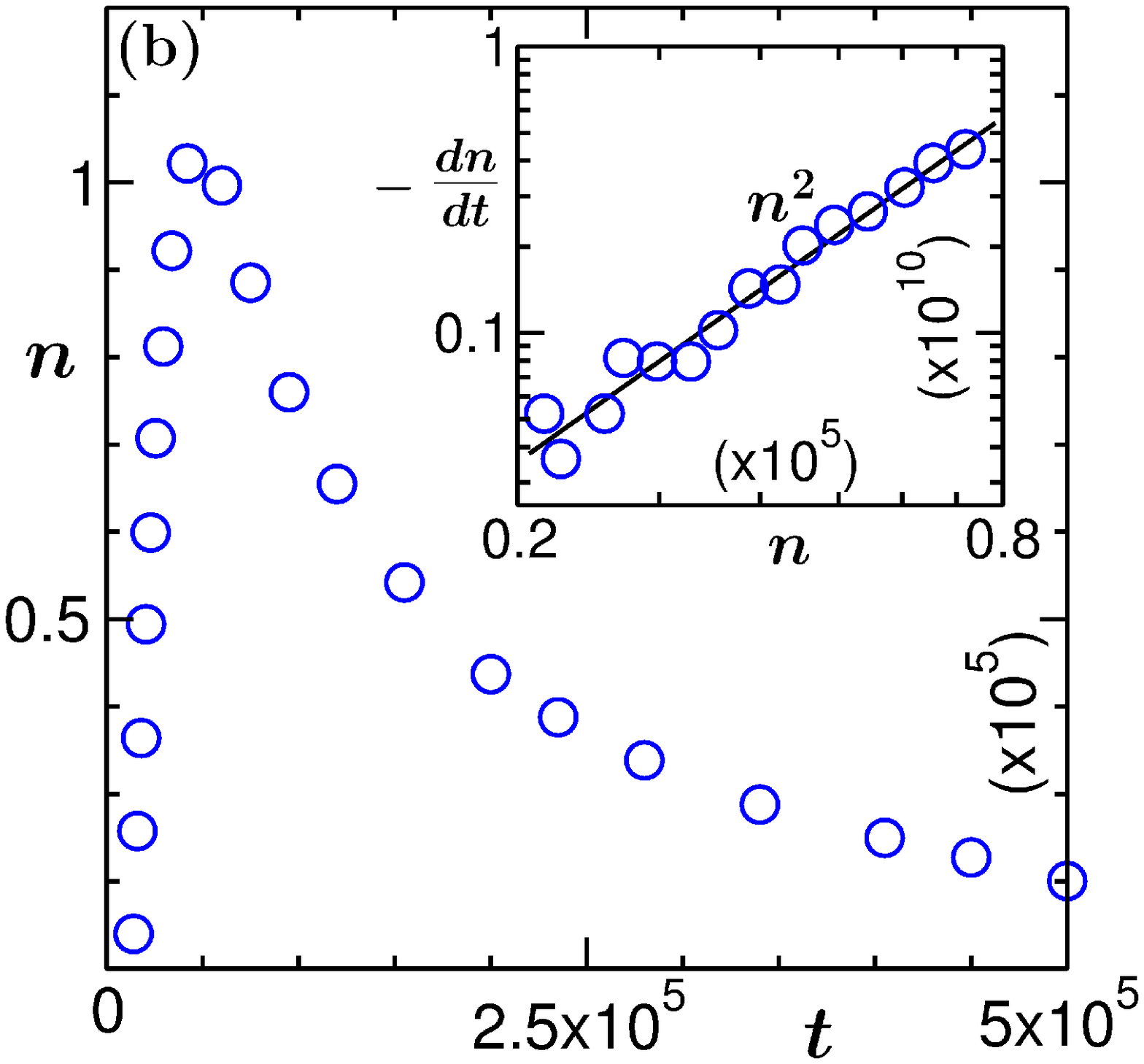}
\caption{\label{fig3_7} (a) The average radius of the liquid droplets is plotted as a function of time, on a log-log scale. 
The solid line represents a power-law growth with exponent $0.5$. The inset shows the evolutions of 
the vapor-phase density and its rate of change. The ordinates have been appropriately multiplied 
($\rho_v \times10$ and $\frac{d\rho_v}{dt}\times 10^7$) for the convenience of presentation. (b) Plot of the droplet density 
as a function of time. In the inset we show negative of $dn/dt$ (for $t>10^5$) as a function of $n$, on a log-log scale. 
The solid line there is a power-law with exponent 2. The axes are multiplied by large numbers.}
\end{figure}
\par
Despite growth via the diffusive motion of the droplets and sticky collisions among them, contribution to the growth due 
to the LS mechanism \cite{chap3_IMLifshitz}, i.e., via evaporation of particles from a smaller droplet and 
their condensation on a bigger one, is still possible. 
To check for that, in Fig. \ref{fig3_6} we show the numbers of particles in a few droplets, over the time scale of typical 
collision interval. These plots convey the message that the sizes of the droplets do not change between collisions. 
Thus, the growth essentially 
occurs via the diffusive droplet coalescence mechanism \cite{chap3_KBinder2,chap3_KBinder3,chap3_EDSiggia}. In that case, we expect a 
power-law growth with $\alpha=0.5$. 
\begin{figure}[h!]
\centering
\includegraphics*[width=0.4\textwidth]{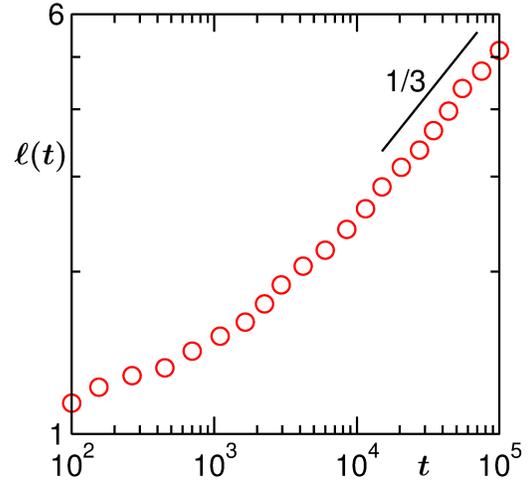}
\caption{\label{fig3_8} Same as Fig. \ref{fig3_7}(a), but the data set was obtained via the application of Andersen thermostat 
in our MD simulations. Unlike the previous results for kinetics (all of which were obtained by using NHT),
these results are presented after averaging over five independent initial configurations with $L=512$. The solid line 
represents a power-law with exponent $1/3$.}
\end{figure}
\par
In Fig. \ref{fig3_7}(a) we show $\ell$ as a function of $t$, on a log-log scale. The late time data appear quite straight, implying power-law.
The solid line there represents the BS growth-law with which the simulation data are very much consistent. 
Slightly faster trend in the simulation data, during an intermediate time regime, can be explained in the following way. Given that 
a perfect linear behavior in the MSD of cluster CMs appears at a rather late time, many collisions, particularly during the 
above mentioned regime, occur while the participating droplets are moving ballistically. This brings a shorter time scale 
in the problem during which the collision partners might not have gained a proper circular shape, from which 
they departed during the previous collisions. Presence of such 
non-circular shape or fractality (see the snapshot at $t=2.5\times10^4$) during ballistic 
aggregation can enhance the growth rate \cite{chap3_JMidya2}. On the other hand,
the slower growth at very early time can be due to the LS 
mechanism \cite{chap3_IMLifshitz}. During this period, the (unequilibrated) vapor phase 
density is rather high, leading to slow movement 
of the droplets. In that case growth can be dominated by evaporation-condensation mechanism.
Nevertheless, we caution, the data from this early period should not be interpreted too seriously, 
for the reason stated above, i.e., the density inside the clusters is still changing, altering the value of 
$\ell$ even if the number of particles inside the droplets remain same. 
To demonstrate this, in the inset of Fig. \ref{fig3_7}(a) we have 
shown the time dependence of the vapor-phase density and its 
time derivative. While it is seen that the variation of this density is negligible in the BS scaling regime, there 
exists significant drop in it before that regime. Such a saturation time scale will be longer as one quenches the systems to 
temperatures closer to the critical value. This is due to the divergences of the (equilibrium) correlation 
length and the time of relaxation over that length scale \cite{chap3_AOnuki,chap3_DPLandau,chap3_VPrivman}. 
In that case, to observe growth purely due to the BS mechanism,  
one needs to consider large enough systems so that the simulations 
can be run for long periods of time without encountering any finite-size effects. Otherwise, 
for the whole period of simulation, one will observe growth with significant contribution coming from the
condensation of particles on a droplet from the vapor phase or from the neighboring droplets.
On the other hand, choice of temperature far below $T_c$ may not allow us to study the kinetics of vapor-liquid transition, 
due to crystallization \cite{chap3_JMidya2}. In the latter situation, the above mentioned fractality of the clusters become very 
prominent and growth may occur via the ballistic aggregation mechanism for the entire period
due to very low density of the vapor phase \cite{chap3_JMidya2}. 
In this case the structural self-similarity is also violated.
\par
While the exercises in Figs. \ref{fig3_5} and \ref{fig3_6} already provide good evidence
that the observation of $\alpha=1/2$ is related to the BS mechanism, we present more results to confirm it further.
In Fig. \ref{fig3_7}(b) we show a plot of the droplet density vs $t$. The increase in $n$, at the initial period, 
is related to nucleation of droplets. The decay at late time should be appropriately analyzed to rule out possibilities other 
than the BS mechanism. In the inset of this figure we show $-dn/dt$ vs $n$, on a log-log scale. The data exhibit power-law, 
the exponent being consistent with $2$. This verifies Eq. (\ref{chap3_bs_law}), the starting point for obtaining $\alpha=1/2$ 
due to the BS mechanism.
\par
Finally, in Fig. \ref{fig3_8} we show a plot of $\ell$ vs $t$ from MD simulations using the Andersen thermostat (AT) \cite{chap3_DFrankel}. 
The AT being a stochastic thermostat, it does not preserve hydrodynamics \cite{chap3_JPHansen}. 
We have checked that for this method the droplets remain quite static. Thus, 
the only possibility of growth is via the LS mechanism. The consistency of the data set with the $t^{1/3}$ behavior, at late time, confirms 
this fact, alongside making sure that the BS mechanism is an hydrodynamic effect.
\section{Summary}
~ We have presented results related to the phase behavior and kinetics for the vapor-liquid transition in a two-dimensional 
Lennard-Jones model. While the phase diagram was obtained via the Monte Carlo simulations 
\cite{chap3_AZPanagiotopoulos,chap3_DPLandau}, 
for the kinetics we have performed molecular dynamics (MD) simulations \cite{chap3_DFrankel} with hydrodynamics preserving 
thermostats. 
Even though MD simulations in the microcanonical ensemble preserves hydrodynamics 
perfectly, simulations in the canonical ensemble become essential 
to study the kinetics of phase separation, particularly for the transitions driven 
by temperature. This is due to the fact that with the increase of domain size, 
as the potential energy of the system decreases, simulations 
in the microcanonical set up will provide continuous increase in the kinetic energy, since the total energy is conserved in this 
ensemble \cite{chap3_DFrankel,chap3_DPLandau}. 
Thus, eventually the system temperature will go above the critical value, discarding the objective. 
\par
We have pointed out the structural difference between the high and the low density quenches. For 
the low density quench we have demonstrated 
the structural self-similarity, identified the growth mechanism and quantified the power-law growth exponent. We have shown that the 
growth essentially occurs due to 
the diffusive motion of the droplets and sticky collisions among them. The identified growth exponent matches 
well with the value predicted by Binder and Stauffer \cite{chap3_KBinder2,chap3_KBinder3,chap3_EDSiggia}, for such a mechanism.
\par
For the lower temperature quenches with similar density we observe interesting disconnected fractal clusters, 
growth of which violate the ``standard'' self-similarity property discussed above \cite{chap3_JMidya2}. 
Growth in this case occurs via the ballistic aggregation mechanism, that provides an 
exponent much higher than the BS value \cite{chap3_JMidya2}.
For the present temperature, we have not observed any such significant long-range order in the droplet phase,
over the time scale of our simulations. We, however, expect the BS mechanism to be valid even for solid clusters as long as
the motion of the clusters is diffusive and the corresponding time scale is comparable to that of the relaxation within the clusters.
For more detailed studies involving various different phases, it will be useful to
obtain a more complete (including the vapor-solid part) and accurate phase diagram.
\par
{\bf Acknowledgment:} 
SKD and JM acknowledge financial supports from the Department of Science and Technology, Government of India. SKD
is also grateful to the Marie Curie Actions plan of the European Union (FP7-PEOPLE-2013-IRSES Grant No. 612707, DIONICOS)
for partial support.
JM is grateful to the University Grants Commission, India, for research fellowship.\\
\\
*das@jncasr.ac.in
\\

\end{document}